%% file: main.tex
\pdfoutput=1

\documentclass[a4paper,12pt]{article}
\usepackage[utf8]{inputenc}
\usepackage[english]{babel}

\usepackage{mathptmx,graphicx,graphics,authblk,url,authblk}
\usepackage[singlespacing]{setspace}
\usepackage[margin=1in]{geometry}
\usepackage[numbers]{natbib}
\usepackage{xcolor,pgfplots,pgfplotstable,tikz}

\title{A Cross-Country Comparison of Crowdworker Motivations}

\author[1,2]{Lisa Posch}
\author[1]{Arnim Bleier}
\author[1]{Fabian Flöck}
\author[1,3]{Markus Strohmaier}
\affil[1]{\textit{Department of Computational Social Science, \protect\\ GESIS -- Leibniz Institute for the Social Sciences, Germany}}
\affil[2]{\textit{Institute of Interactive Systems and Data Science, \protect\\ Graz University of Technology, Austria}}
\affil[3]{\textit{HumTec Institute, \protect\\ RWTH Aachen University, Germany}}

\date{}

\input{macros}

\begin{document}

\interfootnotelinepenalty=10000

\maketitle

\vspace{-2em}

\begin{center}
\newline
\end{center}

\input{introduction}

\input{mcms}

\input{empirical_usa}

\input{differences}

\input{conclusion}

\bibliographystyle{plain}
\bibliography{references}

\end{document}

%% file: macros.tex
\definecolor{cusa}{HTML}{9e0142}
\definecolor{cspain}{HTML}{d53e4f}
\definecolor{cgermany}{HTML}{f46d43}
\definecolor{cbrazil}{HTML}{fdae61}
\definecolor{crussia}{HTML}{fee08b}
\definecolor{cmexico}{HTML}{d1ed45}
\definecolor{cindia}{HTML}{abdda4} 
\definecolor{cindonesia}{HTML}{66c2a5}
\definecolor{cphilipines}{HTML}{3288bd}
\definecolor{cvenezuela}{HTML}{5e4fa2}

\pgfplotsset{compat=1.14}

%% file: introduction.tex
\textbf{Introduction.} 
Crowd employment is a new form of short term employment that has been rapidly becoming a source of income for a vast number of people around the globe. It differs considerably from more traditional forms of work, yet similar ethical and optimization issues arise. One key to tackle such challenges is to understand what motivates the international crowd workforce. In this regard, we study one particularly prevalent type of crowd employment: micro-tasks. So far, the motivation of micro-task crowdworkers has mainly been investigated on Amazon Mechanical Turk (AMT)  (e.g. \cite{kaufmann2011more, naderi2014development, ipeirotis2010demographics}). Consequently, research has focused on workers from the USA and India, which constitute the main population on AMT \cite{Ipeirotis2010analyzing}. 
This work, in contrast, addresses crowd employment as an increasingly international phenomenon by exploring the more workforce-diverse micro-task platform  CrowdFlower\footnote{http://www.crowdflower.com/}. 
Furthermore, questionnaires for measuring motivations should be carefully adapted and tested if they are to be used in these new and culturally diverse work environments, in order to produce interpretable results.
To this end, we developed a novel scale that marks a major step in reliably measuring the motivations driving micro-task participation around the world, and we report on the results of applying this scale, which unveil significant international differences.

%% file: mcms.tex
\textbf{The Multidimensional Crowdworker Motivation Scale.} The Multidimensional Crowdworker Motivation Scale (MCMS) \cite{posch2017measuring} is theoretically grounded in Self-Determination Theory (SDT) and builds mainly on two motivation scales developed for the traditional work context: WEIMS \cite{weims} and MWMS \cite{MWMS}. The stem question of the MCMS is ``Why do you or would you put efforts into doing CrowdFlower tasks?", adapted from MWMS \cite{MWMS} in order to capture both actual and latent motivations. SDT specifies different types of motivation \cite{deci2002overview, gagne2005self}, based on which we adopted the following six constructs for the MCMS: 
\textbf{\emph{Amotivation}} refers the absence of motivation, a state of acting passively or not intending to act all. 
Individuals motivated by \emph{external regulation} act in order to obtain rewards or avoid punishments. We divided this construct into \textbf{\emph{material external regulation}} and \textbf{\emph{social external regulation}}. Another type of motivation is
\textbf{\emph{introjected regulation}}, which aims at the avoidance of guilt or at attaining feelings of worth.
\textbf{\emph{Identified regulation}} refers to a type of motivation where the action is in alignment with the individual's personal goals.
\textbf{\emph{Intrinsic motivation}} is a type of motivation where people are driven by interest and enjoyment. 
The MCMS contains three items per construct, each measured along a 7-point Likert scale.

The MCMS was validated on responses collected from CrowdFlower users of the following ten countries: USA, Spain, Germany, Brazil, Russia, Mexico, India, Indonesia, the Philippines and Venezuela. After removing spam, the respondent sample sizes for the countries ranged from 401 to 722.
For validating the MCMS, we conducted confirmatory factor analysis on the total sample as well as on all country samples individually. The model had a satisfactory fit (CFI between 0.931 and 0.975, RMSEA between 0.038 and 0.056, SRMR between 0.037 and 0.050). In order to assess the cross-country comparability of motivations measured with the MCMS, we conducted measurement invariance tests, which showed that configural, full metric and partial scalar invariance was achieved. 
To the best of our knowledge, the MCMS is the first scale to be validated as an instrument for reliably measuring international crowdworker motivations.

%% file: empirical_usa.tex
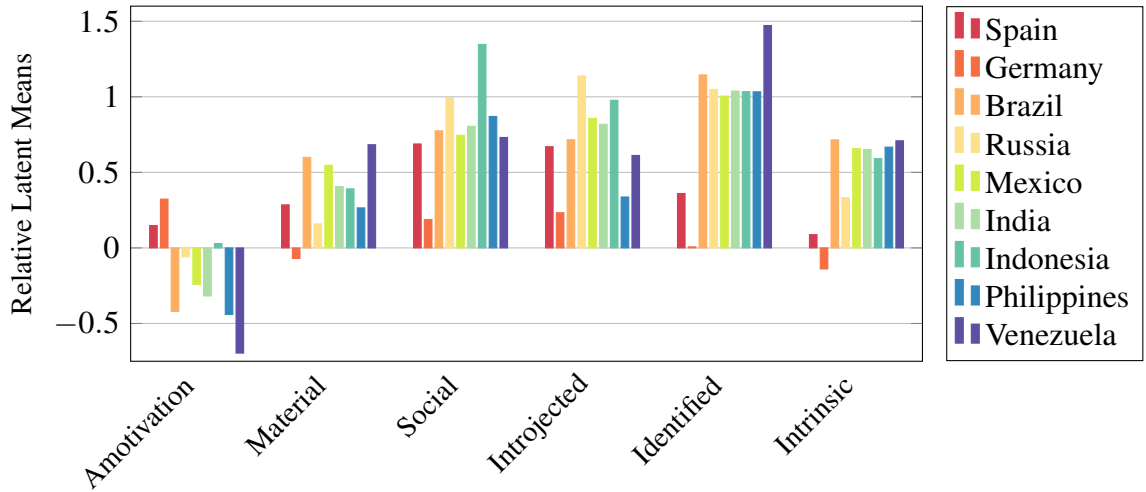
\begin{figure}[t]
	\centering
	\begin{center}
            \input{plot_means_grouped}
	\end{center}	
  	\caption{Country differences in latent means, compared to the USA sample.}
  \centering
  \label{fig:differences}
\end{figure}

\textbf{Results.} As partial scalar invariance was achieved, the analysis of cross-country differences in motivations measured with the MCMS relies on latent means estimated by the model (see Figure \ref{fig:differences}) instead of observed means. In order to estimate country differences in latent constructs, one country has to be chosen as the reference group. For our analysis, we chose the USA sample as the reference group (not shown in Figure \ref{fig:differences}).

\emph{Empirical Construct Means of the Reference Group (USA).} For U.S. crowdworkers, material external regulation was the most important motivational factor (mean 5.86), followed by intrinsic motivation (mean 5.28); this points to an interesting duality of monetary and interest-driven, enjoyment-based motivational influences.
The factor with the third highest mean was identified regulation (mean 3.38), which signifies that putting effort into CrowdFlower tasks is moderately in alignment with American\footnote{We use country demonyms synonymously with the location of workers for better readability.}
crowdworkers' personal goals and objectives.
Amotivation, social external regulation and introjected regulation were the least important motivational factors for crowdworkers in the USA (means 1.91, 1.67 and 1.58, respectively).

%% file: plot_means_grouped.tex
\begin{tikzpicture}
    \begin{axis}[
        width  = 12cm,
        height = 6.29cm,
        major x tick style = transparent,
        ybar=1.2pt,
        bar width=0.1cm,
        ymax=1.6pt,
        ymin=-.75pt,
        ymajorgrids = true,
        ylabel = {Relative Latent Means},
        symbolic x coords={Amotivation,Material,Social,Introjected,Identified,Intrinsic},
        xtick = data,
        ytick = {-0.5,0.0,0.5,1.0,1.5},
       nodes near coords align={vertical},
       x tick label style={rotate=45,anchor=east,font=\small},
       ylabel style={font=\small},
       scaled y ticks = false,
       legend pos=outer north east,
       legend cell align={left},
    ]

        \addplot[style={cspain,fill=cspain,mark=none}]
             coordinates {(Amotivation, 0.148) (Material,0.285) (Social, 0.688)(Introjected, 0.67) (Identified, 0.36)(Intrinsic, 0.088)};
             
        \addplot[style={cgermany,fill=cgermany,mark=none}]
             coordinates {(Amotivation, 0.323 ) (Material,-0.07) (Social, 0.188)(Introjected, 0.233) (Identified, 0.007)(Intrinsic, -0.139)};

        \addplot[style={cbrazil,fill=cbrazil,mark=none}]
             coordinates {(Amotivation,  -0.421) (Material,0.599) (Social, 0.775)(Introjected,0.716) (Identified, 1.145)(Intrinsic, 0.715)};

        \addplot[style={crussia,fill=crussia,mark=none}]
             coordinates {(Amotivation,  -0.059) (Material,0.159 ) (Social,0.994 )(Introjected,1.138 ) (Identified, 1.049)(Intrinsic, 0.332)};
             
        \addplot[style={cmexico,fill=cmexico,mark=none}]
             coordinates {(Amotivation, -0.241 ) (Material,0.547) (Social,0.746 )(Introjected, 0.857) (Identified,1.005  )(Intrinsic, 0.658)};

        \addplot[style={cindia,fill=cindia,mark=none}]
             coordinates {(Amotivation, -0.318 ) (Material,0.406) (Social, 0.805)(Introjected, 0.818) (Identified, 1.038)(Intrinsic, 0.651)};

        \addplot[style={cindonesia,fill=cindonesia,mark=none}]
             coordinates {(Amotivation, 0.029 ) (Material,0.391) (Social, 1.347)(Introjected, 0.977) (Identified,1.034 )(Intrinsic, 0.591)};

        \addplot[style={cphilipines,fill=cphilipines,mark=none}]
             coordinates {(Amotivation, -0.44 ) (Material,0.265) (Social, 0.87)(Introjected, 0.337) (Identified,1.033 )(Intrinsic, 0.667)};

        \addplot[style={cvenezuela,fill=cvenezuela,mark=none}]
             coordinates {(Amotivation,-0.696  ) (Material,0.683) (Social, 0.731)(Introjected, 0.612) (Identified,1.472)(Intrinsic,0.709)};

        \legend{Spain,Germany,Brazil,Russia,Mexico,India,Indonesia,Philippines,Venezuela}
    \end{axis}
\end{tikzpicture}

%% file: differences.tex
\emph{Cross-Country Differences in Motivation.} 
We use estimates of the latent means for comparing crowdworker motivations across countries. Figure \ref{fig:differences} shows the differences in latent means of the different countries, compared to the reference group (USA). The results show that motivations differ significantly between crowdworkers of different countries. 
The largest differences in motivation, compared to workers in the USA, are with countries that are in income groups lower than the USA.
Crowdworkers in Brazil, Mexico, India, the Philippines and Venezuela had a significantly\footnote{Threshold for all reported significances is set at $p < 0.001$, except amotivation in Spain, external social regulation in Germany and external material regulation in Russia ($p < 0.05$).} lower amotivation score than U.S workers, while German and Spanish crowdworkers exhibited a significantly higher level of amotivation. 
Crowdworkers of all countries except Germany reported a significantly higher material external regulation than U.S. workers, which means that workers in these countries are more motivated by material rewards such as money gained.
Social external regulation and introjected regulation was significantly higher in all countries, compared to the USA sample. This means that social punishments or rewards, as well as the avoidance of guilt feelings, are more important motivational factors for workers in countries other than the USA.
Identified regulation was significantly higher in all samples except the German sample. This indicates that for workers of most countries, putting efforts into micro-tasks is more in alignment with achieving personal goals (e.g. career goals) than for U.S. workers.
Finally, all countries except Spain and Germany reported a significantly higher intrinsic motivation than workers in the USA, meaning that workers of most other countries are more driven by interest and enjoyment inherent in the activity.

%% file: conclusion.tex
\textbf{Conclusion.} The results presented in this work constitute a major step towards understanding crowd employment as an international phenomenon. 
Using the MCMS, a reliable instrument for measuring crowdworker motivations, we have presented a first cross-country comparison of crowdworker motivations on the micro-task platform CrowdFlower.
This data provides novel insights regarding wide-ranging national differences of the motivations for participating on such a platform. Enabled by this data, in future work we plan to further investigate the relationship between motivations and economic as well as demographic factors, going beyond the country of residence as an indicator of difference.  
This work is relevant not only for researchers but also for practitioners seeking to harness knowledge about cross-country differences of crowdworker motivations.